\documentclass[twocolumn,superscriptaddress,showpacs,aps,prb]{revtex4}
\usepackage{mathtext}
\usepackage{graphics}
\usepackage[dvips]{epsfig}
\usepackage{amsmath}
\usepackage{slashbox}
\newcounter{fig}

\begin{document}
\title{Thermodynamics of electron-hole liquids in graphene }
\author{L.A. Falkovsky}
\affiliation{L.D. Landau Institute for Theoretical Physics RAS, 119334 Moscow
\\L.F. Verechagin Institute of the High Pressure
Physics  RAS,  142190 Troitsk}
\date{\today}

 \begin{abstract}
The impact of renormalization of the electron spectrum on the chemical potential, heat capacity, and oscillating magnetic moment is  studied. The  cases of low and high temperatures are considered. At low temperatures, doped graphene behaves as the usual Fermi liquids with the power temperature laws for  thermodynamic properties. However, at high temperatures and relatively low carrier concentrations, it exhibits the collective electron-holes features: the chemical potential tends to its value in the undoped case  going with the temperature to the charge neutrality point. Simultaneously, the electron contribution into the heat capacity tends to the constant value, as in  
the case of the Boltzmann statistics.
\pacs{65.80.+n,71.70.Di, 71.18.+y}
\end{abstract}
\maketitle

\section{Introduction}

\begin{figure}[]
\resizebox{.4\textwidth}{!}{\includegraphics{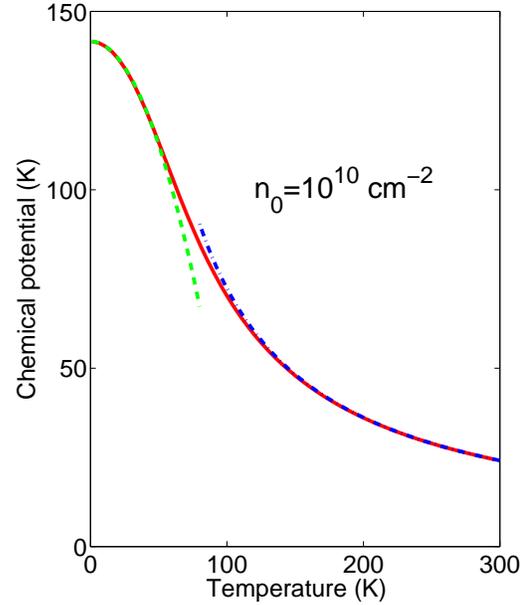}}
\caption{Chemical potential versus temperature for the carrier concentration $10^{10}$ cm$^{-2}$;  the exact solution to Eq. (\ref{num}) is shown by the solid line, the asymptotes  are for  low, Eq. (\ref{mu1}), and high, Eq. (\ref{high}), temperatures (dashed and dashed-dotted lines, correspondingly); the renormalization is not included. }\label{unscr}
\end{figure}

Optic and magneto-optic experiments with graphene layers have been successfully interpreted \cite{NGP} so far in a scheme of massless relativistic particles with a conical energy spectrum 
\begin{equation}
\varepsilon_s(p)=\mp v p
\label{ed}\end{equation}
where $v$ is the constant velocity parameter  in two bands, $s=1,2$, near the K and K' points in the Brillouin zone.
In pure graphene, the chemical potential is situated at the charge neutrality point  $\varepsilon=0$. However, it can have a nonzero value
because of doping or under a gate voltage. Thus, the chemical potential is determined by the total number of carriers (difference of electrons in the upper band and holes in the low band)
 \begin{equation}
 N=4S\int |f\left( \varepsilon -\mu \right) -f\left( \varepsilon +\mu
\right)|\frac{d^{2}\mathbf{p}}{\left( 2\pi \hbar \right) ^{2}}\, ,
\label{num}\end{equation}
where  $f\left( \varepsilon -\mu \right)$ is the Fermi function, $S$ is the surface of the graphene layer, and the factor 4
takes the valley and spin degeneracy into account. The integration is performed over $\varepsilon>0$, the chemical potential  is positive for electrons and negative  for holes. At the fixed $N$,
this condition  determines the dependence  $\mu(T)$, shown in Fig. \ref{unscr}  for a relatively  low electron concentration. 

For the conical spectrum, Eq. (\ref{ed}), the ratio between the kinetic and Coulomb energies has a constant value independently of the carrier concentration and the problem of the phase electron-dielectric transition becomes undefined.
It was recently  discovered \cite{YJB} in studying of the Shubnikov-de Haas oscillations that electron-electron interactions are very important for  low carrier concentrations, $p\rightarrow 0$.
While the electron concentration decreases  from 10$^{12}$ to 10$^9$ cm$^{-2}$, the velocity parameter $v$ grows by three times from its ordinary value 1.05$\times10^8$ cm/s. 
 The logariphmic renormalization of the velocity for the linear electron dispersion was found by Abrikosov and Beneslavsky \cite{AB} in the three-dimensional case and in Refs. \cite{GGV, Mi, BPP, KUP} for two-dimensional graphene.    Notice, that no phase transition  was revealed even at the lowest carrier concentration. We can conclude that   Coulomb interactions do not create any gap  in the spectrum.
 
  The renormalized electron dispersion can be written in the form
\begin{equation}
\varepsilon_s(p)=\mp vp[1+g\ln(p_0/p)]\,,
\label{ren}\end{equation}
where $g=e^2/8\pi \hbar v\epsilon $ is the dimensionless electron-electron interaction  and $\epsilon\simeq 2.5$ describes an effect of a substrate and self-screening in graphene, 
$p_0\simeq 0.5\times10^8$ cm$^{-1}$ is the cutoff parameter \cite{YJB}. In Fig. \ref{scr}, we consider a screening effect on the chemical potential  at the carrier concentration $n_0=10^{12}$ cm$^{-2}$. 

Equation (\ref{ren}) is written in the linear  approximation in $g\ln(p_0/p)<1$. Because the logarithm is assumed to be large, the condition $ g\ll 1$ has to be fulfilled, and we suppose this condition in what followed. In this article, we consider  the impact of the renormalization on  thermodynamic properties of graphene such as the chemical potential, heat capacity, and magnetic moment.   

\section{Temperature dependence of the chemical potential}
For  low  ($\mu\gg T$) and high ($\mu\ll T$) temperatures, the analytical expressions for $\mu(T)$ can be obtained from Eq. (\ref{num}) with the renormalization taken into account.

 For  low temperatures, it is convenient to differentiate Eq. (\ref{num}) with respect the temperature, using
\begin{eqnarray*}
\frac{df\left( \varepsilon -\mu \left( T\right) \right) }{dT}  
=\left[ \frac{ \varepsilon -\mu  }{T}+ \frac{d\mu }{dT}%
 \right] \left[ -\frac{\partial f\left( \varepsilon -\mu \right) }{%
\partial \epsilon }\right]\,.
\end{eqnarray*}
Here, we have a sharp function of $(\varepsilon-\mu)$. Therefore, in the integrand, the momentum
$$p=\varepsilon[1-g\ln(p_0v/\varepsilon)]/v$$
 should be expand near $\varepsilon=\mu$ in  powers of ($\varepsilon-\mu$), which gives  a factor proportional to $T$ after the integration. For instance,  we get in the case of  electron doping
\begin{eqnarray*}
0=\int_{-\infty}^{\infty}\left[ -\frac{\partial f\left( \varepsilon -\mu \right) }{%
\partial \epsilon }\right]\left[\mu\frac{d\mu}{dT}+\frac{(\varepsilon-\mu)^2}{T}\right]\\ \times [1-2g\ln(p_0v/\mu)]d\varepsilon\,,
\end{eqnarray*}
where we do not differentiate the logarithm because of the condition $g\ll 1$.
Integrating, one finds
 \begin{equation}
\frac{d\mu}{dT}=-\frac{\pi^2}{3\varepsilon_F}T\,,
\label{mu1}\end{equation}
where we denote $\varepsilon_F\equiv\mu(T=0)$, positive for electrons and negative for holes. Let us notice that this is the known temperature dependence of the chemical potential in the degenerate Fermi system at low temperatures.    We emphasize that the Fermi energy $\varepsilon_F$ is determined indeed by the carrier concentration
\begin{equation}
n_0=\frac{p_F^2}{\pi\hbar^2}=\frac{1}{\pi}\left(\frac{\varepsilon_F}{\hbar v}\right)^2[1-2g\ln(p_0v/|\varepsilon_F|)]\,,
\label{mu3}\end{equation}
which introduces the renormalization in Eq. (\ref{mu1}) by means of $\varepsilon_F$.

\begin{figure}[b]
\resizebox{.4\textwidth}{!}{\includegraphics{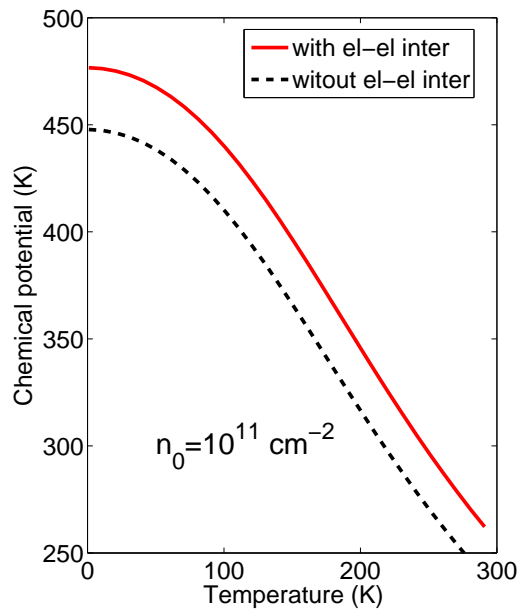}}
\caption{Chemical potential versus temperature for the carrier concentration $10^{12}$ cm$^{-2}$; the renormalization is included (solid line), the chempotential for  noninteracting electrons  is shown in the dashed  lines; the cutoff parameter $p_0=0.5\times10^8$ cm$^{-1}$, the dielectric constant $\epsilon= 2.5$. }\label{scr}
\end{figure}

For  high temperatures, we can expand  the integrand in Eq.
(\ref{num}) in $\mu $. Introducing the new variable $x=\varepsilon/2T$, we get the integral
\begin{eqnarray*}
N=\frac{4|\mu| ST}{\pi(\hbar v)^2 }\int_{0}^{\infty}\frac{1-2g\ln(p_0v/2Tx)}{\cosh^2 x}xdx\,,
\end{eqnarray*}
which gives the chemical potential
\begin{equation}|\mu|=\frac{\pi}{4\ln{2}}\frac{ n_0 (\hbar v)^2}{ T}[1+2g\ln(p_0v/2T)]\,.\label{high}\end{equation}
We see the inverse temperature dependence of the chemical potential, as a collective effect in  electron-hole liquids. The renormalization  term, correcting the temperature dependence, is presented here explicitly and illustrated in Fig. \ref{scr}.

\section{Heat capacity}
Now we consider the electron contribution in the heat capacity. The  energy of carriers
\begin{equation}
E=4S\int_0^{\infty} \varepsilon |f\left( \varepsilon -\mu \right) -f\left( \varepsilon +\mu
\right) |\frac{d^{2}\mathbf{p}}{\left( 2\pi \hbar \right) ^{2}} 
\label{eneg}\end{equation}%
differs from the carrier concentrations, Eq. (\ref{num}), only by the additional factor $\varepsilon$ in the integrand. Therefore, we can follow the same procedure.

For low temperatures, $T\ll \varepsilon_F$,  the carrier heat capacity
 in the case of electron doping writes as
\begin{eqnarray*}
C_{S}^{\left( e\right) } =\frac{2S}{\pi \left( \hbar v\right) ^{2}} \int_{-\infty}^{\infty }\left[ \mu ^{2} \frac{
d\mu }{dT}+2\mu \frac{\left( \varepsilon -\mu \right)^2 }{T}
\right] \\ \times \left[ -\frac{\partial f\left( \varepsilon -\mu
\right) }{\partial \varepsilon }\right][1-2g\ln(p_0v/\mu)] d\varepsilon\\ =\frac{2S}{\pi \left( \hbar v\right) ^{2}}\left[ \mu ^{2}\frac{d\mu }{dT}+\frac{2\pi^2}{3}\mu T\right][1-2g\ln(p_0v/\mu)]\,. 
\end{eqnarray*}%
Using Eq. (\ref{mu1}), we have 
\[
C_{S}^{\left( e\right) }=\frac{2\pi S|\varepsilon_F |}{3\left( \hbar v\right) ^{2}}T [1-2g\ln(p_0v/|\varepsilon_F|)]
\]
 in both cases of the electron or hole doping.
 
 For high temperatures, $T\gg \mu$, one can perform the expansion of the energy, Eq. (\ref{eneg}), in the first order of $\mu $%
\begin{eqnarray*}
E&=&\frac{4S|\mu |}{\pi \left( \hbar v \right) ^{2}}\int_0^{\infty} \varepsilon ^{2}\left( -%
\frac{\partial f\left( \varepsilon \right) }{\partial \varepsilon }\right)
[1-2g\ln(p_0v/\varepsilon)]d\varepsilon\\& =&%
\frac{2\pi S |\mu |}{3\left( \hbar v \right) ^{2}}T^{2}[1-2g\ln(p_0v/2T)]\,.
\end{eqnarray*}%
 Using Eq. (\ref{high}), we find
\begin{eqnarray*}
E =\frac{\pi ^{2}}{6\ln 2}NT\quad \text{and} \quad
C_S^{(e)}=\frac{\pi ^{2}}{6\ln 2}N\,. \end{eqnarray*}%
Finally,%
\begin{equation}
C_S^{(e)}=\frac{\pi^2 }{3}N \left\{ 
\begin{tabular}{l}
$2\frac{T}{|\varepsilon_{F}|}\,,\quad T\ll |\mu|$ \\ 
$\frac{1}{2\ln 2}\,,\quad T\gg |\mu| $%
\end{tabular}%
\right. 
\label{cap}\end{equation}
Thus, we see that the renormalization modifies the heat capacity  at low temperatures, i.e., in the degenerate statistics. At high temperatures, the heat capacity possesses the constant value and does not reveal any renormalization at least to a first approximation in $g\ln(p_0/p)$.  

\section{Magnetic susceptibility}
Magnetic susceptibility is determined by the dependence of the thermodynamic potential on the magnetic field 
\begin{eqnarray*}
\Omega(B)=-\frac{2eBT S}{\pi\hbar c}\sum_{n,s}\ln\left(1+e^{\frac{\mu-\varepsilon_{sn}}{T}}\right)
\end{eqnarray*}
in terms of the electron dispersion $\varepsilon_{sn}$ for two bands $s=1,2$ with the Landau number $ n=0,1,2..$ We neglect the spin splitting of the levels in comparison with the large Landau splitting in graphene. 

Oscillations of the magnetic moment in the semi-classical region can be found applying the 
Poison formula  to the thermodynamic potential
\begin{eqnarray*}
\Omega(B)&=&- \frac{2eBT S}{\pi\hbar c}\sum_{k\neq0}\int_0^{\infty}\left\{
\ln\left(1+e^{\frac{\mu-\varepsilon}{T}}\right)\right.\\&+&\left.
\ln\left(1+e^{\frac{\mu+\varepsilon}{T}}\right)\right\}e^{2\pi ikn} dn\,,
\end{eqnarray*}
where the contributions of two bands are written explicitly.
The integraton by parts gives
 \begin{equation}
\Omega(B)=-\frac{eBT S}{\pi^2\hbar c}\sum_{k\neq0}\int_0^{\infty}\frac{1}{ik}[
f(\varepsilon-\mu)-f(-\varepsilon-\mu)]e^{2\pi ikn}d\varepsilon\,.
\label{sm}\end{equation}
For the semi-classical region, we  use the quantization rule in the Bohr--Zommerfeld form
\[2\pi n=\frac{cA(\varepsilon)}{e\hbar B}\]
with the aria enclosed by the electron trajectory for the energy $\varepsilon$ in the momentum space   \[A(\varepsilon)=\pi\left(\frac{\varepsilon}{v}\right)^2(1-2g\ln(p_0v/\varepsilon)\]
according to Eq. (\ref{ren}).

The main contribution in the integral (\ref{sm}) comes from the vicinity of the point $\varepsilon=\pm\mu(T=0)=\pm\varepsilon_F$ for the positive and negative $\varepsilon_F$, correspondingly. Expanding the exponent in the integrand near that points and integrating, one finds
\begin{eqnarray*}
\Omega(B)=\frac{2 eBTS}{\pi\hbar c}\sum_{k\ne 0}\frac{1}{k}
\frac{\sin[kcA(\varepsilon_F)/e\hbar B]}{\sinh(2\pi^2 kc|m(\varepsilon_F)| T/e\hbar B)}\,,
\end{eqnarray*}
where $m(\varepsilon)=\frac{1}{2\pi}\frac{dA(\varepsilon)}{d\varepsilon}$ is the cyclotron mass. 
In calculating of the magnetic moment we can derivative only the rapid factor in the argument of $\sin$ with respect $B$: 
\begin{eqnarray}\label{sus}
\tilde{M}(B)=\frac{2\pi n_0 S T}{ B} \sum_{k\ne 0}
\frac{\cos[kcA(\varepsilon_F)/e\hbar B]}{\sinh[2\pi^2 kc|m(\varepsilon_F)| T/e\hbar B]}\,,
\end{eqnarray}
where the carrier concentration $n_0= A(\varepsilon_F)/(\pi \hbar)^2$.
This is the standard Lifshiz-Kosevich formula used in Ref. \cite{YJB} for the interpretation of experimental data concerning the velocity renormalization. There are  two important features: first, the aria $A(\varepsilon_F)$  and the effective mass $m(\varepsilon_F)$ should be taken at  the renormalized  Fermi energy corresponding to the carrier concentration and, second, the factor in front of the sum differs from the 3d case since the integration over $p_z$ is absent now.  

It is interesting to compare the amplitude of oscillations with the monotonic part of the magnetic moment, Ref.  \cite{MC,SD,OK},
\begin{eqnarray*}
M_0=\frac{-S}{6\pi }\left(\frac{ev}{c}\right)^2\frac{B}{T\cosh^2(\mu/2T)}\,.
\end{eqnarray*}
Thus, we see that the ration of the oscillating and monotonic parts of 
the magnetic moment has the order
\[|\tilde{M}/M_0|\sim 12\pi n_0\left(\frac{cT}{evB}\right)^2
\frac{\cosh^2(\varepsilon_F/2T)}{\sinh[2\pi^2 c|m(\varepsilon_F)| T/e\hbar B]}\,.\]
To observe the oscillations, the argument of $\sinh$ has to be small or at least on the order of unity. Then, the monotonic part of the magnetic moment can be observable only at relatively high temperatures, $|M_0/\tilde{M}|\sim \exp(-\varepsilon_F/T)$.
\section{Conclusions}

It should be emphasize that  such a transport property as the electronic conductivity is not sensitive to the electron-electron interaction since the conductivity  does not depend indeed on the velocity parameter $v$. 
The renormalization of the electron spectrum due to Coulomb interactions in graphene is noticeable in thermodynamic properties especially at low temperatures and  for small carrier concentrations $n_0< 10^{10}$ cm$^{-2}$, as can be seen from Eqs. (\ref{high}), (\ref{cap}), and (\ref{sus}).
However, the interesting temperature dependences $\mu\sim T^{-1}$ for the chemical potential at high temperatures $T\gg \mu$, appears independently  of  electron-electron interactions. The detection of the renormalization requires the high accuracy in experiments because the renormalization can be concealed by increasing of the velocity parameter $v$ [see Eq. (\ref{mu3}) and Fig. \ref{scr}].

\acknowledgments

We gratefully acknowledge Andrey Varlamov for useful discussions. This work was supported by the Russian Foundation for Basic
Research (grant No. 13-02-00244A) and the SIMTECH Program, New Centure of Superconductivity: Ideas, Materials and Technologies (grant No. 246937).


\begin{thebibliography}{99}
\bibitem{NGP} A.H. Castro Neto, F. Guinea, N.M.R. Peres, K.S. Novoselov, A.K. Geim, Rev. Mod. Phys. {\textbf 81}, 109 (2009).

\bibitem{YJB} D.C. Elias et al, Nat. Phys. {\textbf 7}, 701 (2011); G.L. Yu et al, arXiv:1302.3967.

\bibitem{AB} A.A. Abrikosov, S.D. Beneslavsky, Sov. Phys. JETP {\textbf 32}, 699 (1971).

\bibitem{GGV} J. Gonzalez, F. Guinea, M.A.H. Vozmediano, Nucl. Phys. B  \textbf{424}, 595 (1994); J. Gonzalez, F. Guinea, M.A.H. Vozmediano,  Phys. Rev B  \textbf{59}, 2474 (1999).

\bibitem{Mi} E.G. Mishchenko, Phys. Rev. Letts. \textbf{98}, 216801 (2007).

\bibitem{BPP} Y. Barlas, T. Pereg-Barnea, M. Polini, R. Asgari, A.H. MacDonald, Phys. Rev. Letts. \textbf{98}, 236601 (2007).

\bibitem{KUP} V.N. Kotov, B. Uchoa, V.M. Pereira, F. Guinea, A.H. Castro Neto, Rev. Mod. Phys. {\textbf 84}, 1067 (2012).

\bibitem{MC} J.W. McClure, Phys. Rev.  \textbf{104}, 666 (1956).

\bibitem{SD} S.A. Safran, F.J DiSalvo, Phys. Rev. B \textbf{20}, 4889 (1979).

\bibitem{OK} Y. Ominato, M. Koshino, arXiv:1301.5440.

\end{thebibliography}
\end{document}